# MDA based-approach for UML Models Complete Comparison

Samia Benabdellah Chaouni[1], Mounia Fredj[2] and Salma Mouline[3]

[1] ENSIAS, Mohammed V Souissi University
Rabat, Av. Mohammed Ben Abdallah Regragui, Madinat Al Irfane, BP 713, Agdal, Morocco
*bcsamia@gmail.com*

[2] ENSIAS, Mohammed V Souissi University
Rabat, Av. Mohammed Ben Abdallah Regragui, Madinat Al Irfane, BP 713, Agdal, Morocco
*fredj@ensias.ma*

[3] Faculty of Sciences, University Department, Mohammed V Souissi University
Rabat, 4 Av.Ibn Battouta B.P. 1014 RP, Morocco
*mouline@fsr.ac.ma*

**Abstract**
If a modeling task is distributed, it will frequently be necessary to integrate models developed by different team members. Problems occur in the models integration step and particularly, in the comparison phase of the integration. This issue had been discussed in several domains and various models. However, previous approaches have not correctly handled the semantic comparison. In the current paper, we provide a MDA-based approach for models comparison which aims at comparing UML models. We develop an hybrid approach which takes into account syntactic, semantic and structural comparison aspects. For this purpose, we use the domain ontology as well as other resources such as dictionaries. We propose a decision support system which permits the user to validate (or not) correspondences extracted in the comparison phase. For implementation, we propose an extension of the generic correspondence metamodel AMW in order to transform UML models to the correspondence model.

## 1. Introduction

In a context of increasing complexity of software systems, the collective modeling (realized by different teams) becomes a major activity of development process, particularly in the design phase, where several business models are produced and must be integrated to obtain a coherent overall business model of the system. The goal is to integrate these models easily and efficiently.
We were interested in the integration of UML models and more specifically in UML class diagrams. From the review of the existing work, it appears that semantic integration is a crucial problem, which has not fully been addressed. In this paper, we focus on models comparison, which is the first stage of the integration process.

We propose a models transformation-based hybrid approach which compares models on syntactical, semantical and structural aspects. For this purpose, we used several resources, of which domain ontology.
This article is organized as follows: Section 2 is an introduction to the general approach of models integration. We mention in section 3 previous work relevant to this issue and their limitations. We present in section 4 the MDA (Model Driven Architecture), the concepts of models, models transformation and ontologies. Our proposal (architecture, comparison rules, comparison by models transformation and correspondence metamodel) is developed in section 5. The reader is provided with a case study in section 6. Finally, our implementation and some research perspectives are developed in the conclusion section.

## 2. Integration of models

The integration is defined as the combination of components in such a way as to form a new set constituting a unit for creating synergy [1]. Existing research [2] [3] has shown that models integration process involves two steps: 1) **the comparison step** is based on a set of rules called correspondence rules (also called comparison rules, mapping rules or matching rules) which identify the correspondence between elements of models (correspondences created during this step are stored in a separate model called **correspondence model** or mapping model); 2) **the integration step** integrates models mapped in the previous step. The integration strategy relies on rules that define which elements will appear in the result model and how elements will be organized in the result model. These rules are i) rules for merging the matched elements (merging rules), and ii) rules for incorporating elements that do not belong to the mapping model (integration rules).

## 3. Existing work

Several studies have proposed models comparison. The authors of [4], [5] and [6] provided a comparison of metamodel independent models. Databases comparison has been treated in [7]. In [8], the authors provided a comparison of aspect oriented UML models. In [9], a comparison of views models is proposed. [10] developed a method to compare UML class diagrams. The specification of UML [11] defines the comparison of packages.

We found different approaches of models comparison:

- Syntactic approaches: they compare the letters of strings of models elements.

- Semantic approaches: they compare the meaning associated with the compared items.

- Local structural approaches: they compare the components of the elements. For example, the comparison of local structure of two classes corresponds to the comparison of their attributes and operations.

- Global structural approaches: they compare the elements that are related to the elements to compare. For example, the comparison of global structure of two relations corresponds to the comparison of the classes connected by these relations.

- Hybrid approaches: they combine two, three or four types of comparison (syntactic, semantic, global structure and local structure).

The diagram below displays a synthesis of these works. References of the approaches are shown on the horizontal axe. The existing types of comparison are provided on the vertical axe. Points show which type of comparison is used by the approach.

Let M1 and M2 be two models to compare. Most approaches compare models elements syntactically. However, they only test the identity of elements. [7] detects also other correspondences such as abbreviation (e.g. "Qty" in M1 and "Quantity" in M2) and acronym (e.g. "UOM" in M1 and "UnitOfMeasure" in M2). Moreover, most approaches compare local structure and/or global structure of models elements. Finally, all of these works do not take into account the semantic aspects and are limited to detection of synonyms (e.g. "Book" in M1 and "Work" in M2) and homonyms (e.g. two classes "Family" (products) and "Family" (people)).

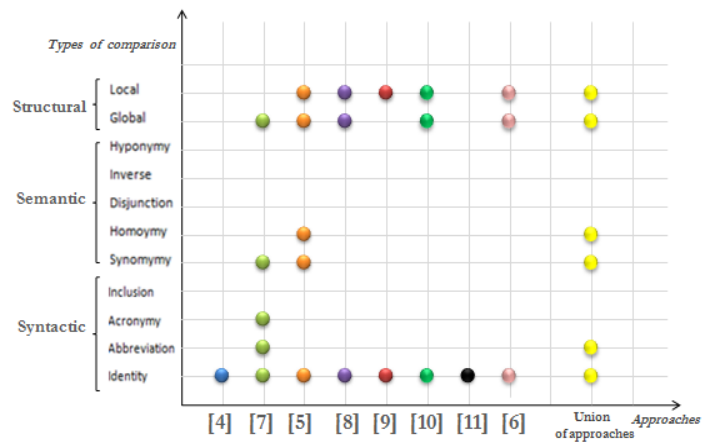

Fig. 1. Comparison diagram of the existing work

Our review shows that existing work does not detect semantic mappings such as disjunction (e.g. two boolean attributes "Single" and "Married") and reverse (e.g. the relationship "Buy" is the inverse of "BoughtBy" relationship). Syntactic correspondences such as inclusion syntactic (e.g. "Student" and "Students") are not either detected. These limits are shown in the diagram by a discontinuity of points in each approach. One may also emphasize that approaches are complementary, even though their union does not cover all types of comparison and does not detect all matches (note that in the diagram, even the union of approaches is a discontinuity of points).

Therefore, our goal is to provide a hybrid approach incorporating syntactic, structural and semantic aspects in order to detect any mapping or correspondence.

## 4. MDA and Ontologies

### 4.1 MDA

MDA is a software development lifecycle process that has been created by the Object Management Group (OMG) and was introduced in 2001 [12]. The main idea behind MDA is to use models as core development artifacts and thus be able to separate platform specific data from the software development process. Developing applications without platform specific terms makes it easier and less costly to port them to different platforms [12].
The two main artifacts of MDA are models and models transformation.

### 4.1.1 OMG's 4-layer metamodeling Architecture

To organize and structure these models, the OMG has defined an architecture called "four-layered architecture". It is an architectural framework for models, metamodels and meta-metamodels. The layers in the four-layered architecture are called M0, M1, M2, and M3. Every layer is an instance of the layer above except for layer M3 which is specified reflexively and therefore does not need layers above.

M0 layer : is the running system in which the actual instances exist. This layer holds the user data, and the actual object that software is designed to manipulate [13,14]. "Layer M0 specifies user objects that are instances of the UML user model classes" [15].

M1 layer : The elements of the M1 layer are models. An example would be a UML model of a software system. M1 layer is a model of the M0 layer user data [13, 14].

M2 layer : holds a model of the information at M1. As it is a model of a model, it is often referred to as a metamodel [14].

M3 layer : defines a model of the information at layer M2, and therefore is often called the meta-metamodel. MOF is the standard for defining the layer M3 elements [14]. "Layer M3 specifies meta-metaclasses for the UML metamodel" [15].

### 4.4.2 Models and Class diagram

*"A model is a simplification of a system built with an intended goal in mind. The model should be able to answer questions in place of the actual system"*[16].

We apply our approach on two simplified models of class diagrams. A class diagram consists of classes, which can be connected pairwise by association or inheritance edges. Furthermore a class can have attributes, described by a name and a type, and operations consisting of a name, a return type and optional ordered parameters described by name and type. Association edges have a name and cardinalities for both anchor points.

### 4.4.3 Models transformations

The MDA guide [12] defines a model transformation as "the process of converting one model to another model of the same system". Kleppe et al. [13] defines a *transformation* as the automatic generation of a target model from a source model, according to a transformation definition. A *transformation definition* is a set of transformation rules that together describe how a model in the source language can be transformed to a model in the target language. A *transformation rule* is a description of how one or more constructs in the source language can be transformed to one or more constructs in the target language.

Rephrasing these definitions by considering figure 2, a model transformation take as input a model conforming to a given source metamodel and a target metamodel. The transformation model contain rules transformation written in transformation language conforming to a metamodel as ATL (Atlas Transformation Language) [17], or QVT [18] (Query View Transform), an OMG's standard, to produce as output another model conforming to a target metamodel.

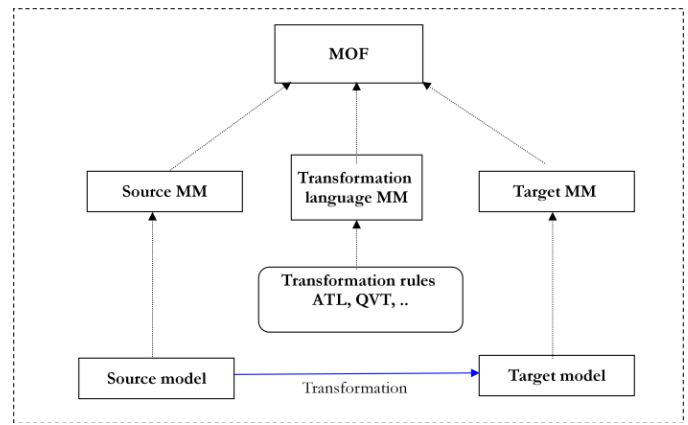

Fig. 2 Models transformation [12, 13]

### 4.2 Ontology

Ontologies are introduced as an "explicit specification of a conceptualization" [19]. Domain ontologies are ontologies which are built on a particular knowledge domain. There are many domain ontologies such as MENELAS (medical domain) [20] and TOVE (business management domain) [21]. The domain ontology is a semantically rich model (it can express equivalence, inverse, disjunction, symmetry, transitivity, etc.), and is defined as an exhaustive list of concepts and relations between these concepts describing a particular domain (Medicine, Business, Library, Restaurants, etc.).

## 5. Our proposal of models comparison

### 5.1 COM$^2$Model system

#### 5.1.1 Architecture

Our goal is to provide a semantic comparison approach integrating syntactic and structural aspects as well. We put forward in [22] a system called COM$^2$Model (Complete Comparison of Models) that takes two models as input and gives correspondence model as output (figure 3). COM$^2$Model is syntactic, semantic and structural rule-based. It detects mappings between models elements. We used strategies based on semantic properties to take into account the semantic aspect. Therefore, our system refers to a domain ontology that will enable to provide semantic relevant information and decision-making during the comparison. We use OWL (Ontology Web language) ontology because it is a W3C recommendation[1], and the metamodel OWL was defined by Ontology Definition Metamodel specification [23] of OMG[2].

Our system is also based on other resources to complete syntactic comparison. We use an acronym dictionary[3], an abbreviation dictionary[4], and a dictionary of synonyms as WordNet[5]. In our approach, we consider that the domain ontology and the other resources exist. We provide a system for decision support. Our system allows the user to validate or to delete mappings that are automatically created.

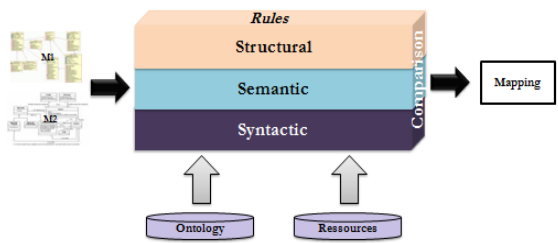

Fig.3. COM2Models architecture Comparison

Our comparison process starts with the syntactical and semantical comparison of elements (first classes, second attributes, third operations and fourth relations). Next, it compares elements (in the same order as just described) in global structures and in local structures.

### 5.1.2 Comparison rules

We provided a first version of rules comparison in informal (natural) language in [24] and an improved version applied to a case study in [25]. To specify the language for expressing these rules, we proposed a metamodel in [22] called Comparison rules metamodel. The formal comparison rules are expressed in [26].

We proposed the following rules:

*Syntactic rules*: identify syntactic identity of two elements (classes, attributes, operations, and relations), syntactic inclusion of two elements, acronyms equivalence of two elements, abbreviation equivalence of two elements and syntactic equivalence of two elements.

---

[1] www.w3c.org
[2] www.omg.org
[3] http://acronymes.info/
[4] http://theleme.enc.sorbonne.fr/dico.php
[5] http://wordnet.princeton.edu/

Example of syntactic rule expressed in informal language: A first element is syntactically included in a second element if the name of the first element appended to a prefix and (or) a suffix gives the name of the second element.

*Semantic rules:* identify synonymy of two elements, equivalence (ontology) of equivalence two classes, equivalence (ontology) of two relations, inverse between two relations, disjunction of two attributes, semantic equivalence of two classes, semantic equivalence of two attributes and semantic equivalence of two relations.

Example of semantic rule expressed in informal language: one relation is inverse of another one, if the two relations are inverse in the ontology.

*Global structural rules*: identify global structure equivalence of two classes, global structure equivalence of two relations, global structure equivalence of two attributes and global structure equivalence of two operations.

Example of g*lobal structural* rule expressed in informal language : Two relations $R_i$ and $R_j$ are equivalent in global structure if: [They link two classes syntactically or semantically equivalent] Or [There are two classes $C_k$ and $C_m$ and there is $C_o$ class such as $C_o$ is the superclass of $C_k$ and $R_i$ links $C_o$ and $C_m$ and there are two classes $C_l, C_n$ such as $R_j$ links them and $C_k$ and $C_l$ are syntactically or semantically equivalent and $C_m$ and $C_n$ are syntactically or semantically equivalent].

*Local structural rules:* identify local structure equivalence of two classes, local structure equivalence of two relations, local structure equivalence of two attributes and local structure equivalence of two operations.

Example of l*ocal structural* rule expressed in informal language: Two classes are equivalent in local structure if their attributes and operations are syntactically or semantically equivalent.

*Equivalence level:* Once these rules are applied to the elements of models, we assign a level of equivalence: i) sure mapping (figure 11), ii) moderately sure mapping (figure 12) and iii) improbable mapping (figure 13). It will help the user to decide to validate or to delete the mapping automatically create by our system.

For instance, if two classes are syntactically or semantically equivalent, and equivalent in global structure (the classes and relations that surround them are equivalents), and equivalent in local structure (the two classes have the same attributes and operations), then it is sure that they are equivalent (*sure mapping)*. So, the user will necessarily validate the mapping.

However, if two classes are equivalent in local structure (the two classes have the same attributes and operations), and are not equivalent syntactically and semantically, and are not equivalent in global structure, then it is sure that they are not equivalent (*improbable mapping*). There is a high probability that the user deletes the mapping.

As shown in [26], we create rules of classes equivalent levels: 1 (L1: improbable mapping), 2 (L2: moderately sure mapping), 3 (L3: sure mapping) and 4 (L4: sure hyponymy relation i.e. a relation that links a superclass to a subclass). We also create rules of attributes / operations / relations equivalent levels: 1 (L1: moderately sure mapping) and 2 (L2: sure mapping), and rule of generalization relations equivalent level 1 (L1: sure mapping). If classes are semantically or syntactically equivalent and are not equivalent in local or global structure, we can talk about homonymy; and if attributes / operations / relations are syntactically or semantically equivalent and non equivalent in global structure then we can talk about homonyms.

## 5.2 MDA-based models complete comparison

### 5.2.1 Comparison by models transformation

The models comparison clearly involves a models transformation because from two models, it creates a correspondence model that contains the correspondence relationships between models elements [27]. In our work (figure 4), models transformation takes as input two UML models conforming to UML metamodel and a correspondence metamodel (presented in the next section). Transformation rules are comparison rules proposed in [26]. The transformation produces as output a correspondence model conforming to the correspondence metamodel.

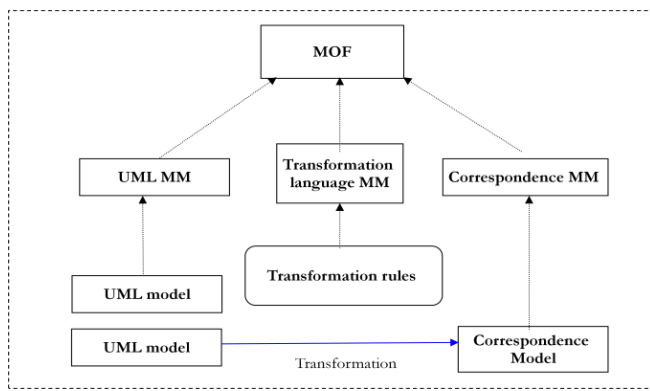

Fig.4 Transformation of UML models into correspondence model

### 5.2.2 Correspondence metamodel

In our work, we suggest using relationships to capture correspondences between models elements. These relationships are stored in a model called "correspondence model", that will be used as input in the integration step.

The correspondence model is conforming to a metamodel that defines the structure of correspondence models. The semantic of correspondence relations mainly depend on application scenario.. The correspondence metamodel contains constructs to define relations between models elements.

To define a correspondence metamodel which satisfies our approach (complete comparison: syntactic, semantic and structural), we extended the generic metamodel Atlas Model Weaver (AMW) proposed in [29]. AMW metamodel allows to make extensions by defining new relation types, according to the application domain.

The ATLAS team, as part of the AMMA project (ATLAS Model Management Architecture), proposed AMW. This metamodel is specifically made to create links or relationships between models elements, through an operation called "weaving". These links are stored in a model conforming to a metamodel named "weaving metamodel", which specifies the semantics of the links. Therefore, AMW provides a set of models called weaving models which thereafter can be used by a transformation tool. Therefore, our metamodel (figure 5) allows defining the concepts needed to manage both existing relationships and new types of links.

The existing AMW metamodel elements are:
- *Element*: it is the base element of all metamodel elements. All the others elements extend it. It has two attributes: *name* and *description*.
- *WModel* is the root element. It is composed of the weaving elements.
- *WLink*: represents the link between the models elements. The reference *end* enables linking between arbitrary numbers of elements. Weaving links can also relate with other weaving links to create a containment relation. This element should be extended to add different linking semantics to the weaving metamodel.

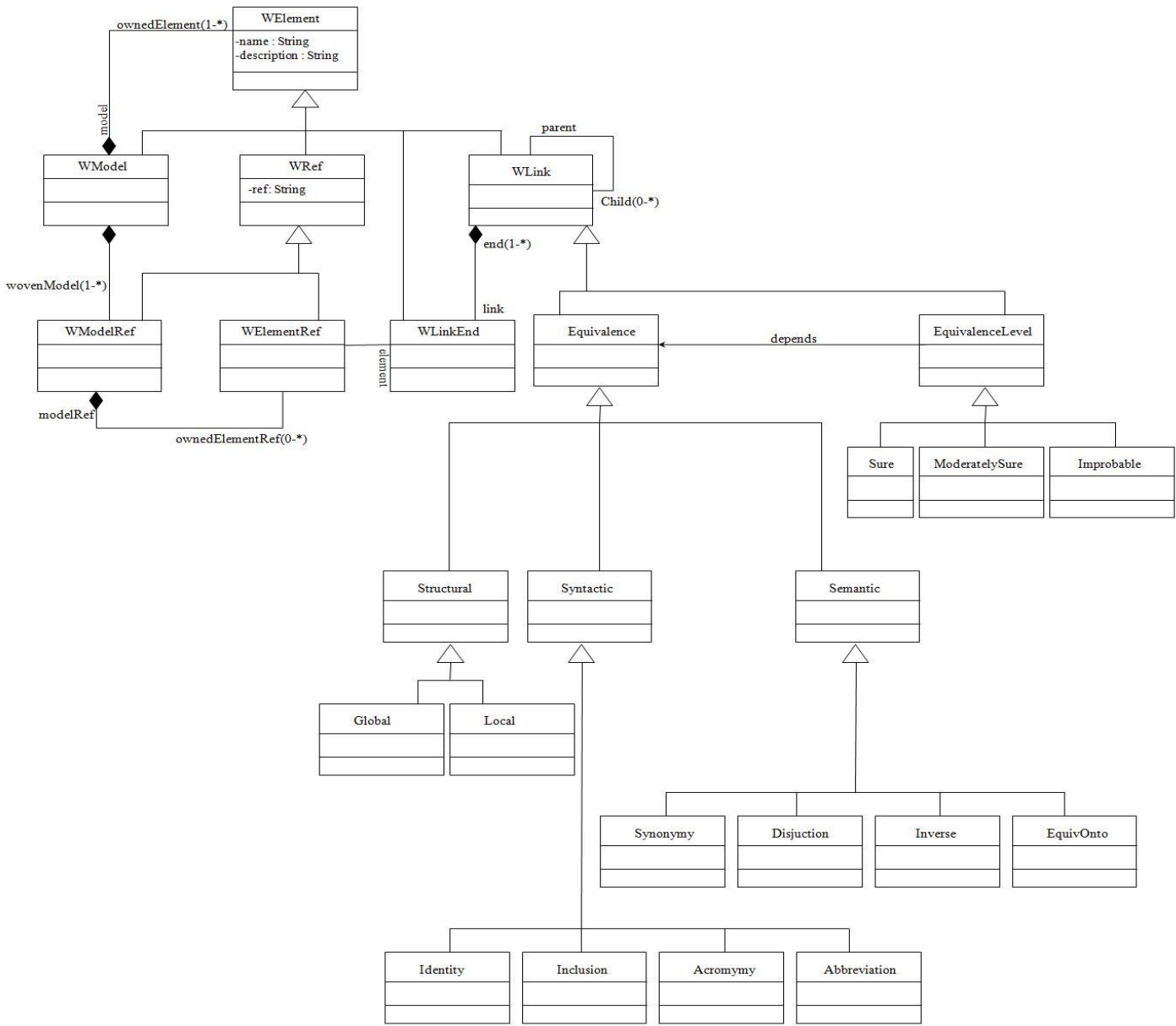

Fig. 5 Proposition of the correspondence metamodel

- *WRef*: abstract class that represents the references.
- *WElementRef* has an identifier (ID) that points to the elements of the input models.
- *WLinkEnd* indicates the extremity of a link, referencing the woven model elements. i.e. indicates the type of elements that are be composed.
- *WModelRef* is similar to *WElementRef* element, but it references an entire model.

Below we present the extension semantic of AWL metamodel, i.e. the new types of links between models elements:
- *Equivalence*: represents the equivalence link between two equivalent elements.
- *Identity*: indicates that the two elements related to each other by this relation are syntactically identical.
- *Acronym*: indicates that the two elements related to each other by this relation are acronyms.
- *Inclusion* : indicates that one of the two elements related to each other by this relation is syntactically included in the other element.
- *Abbreviation:* indicates that one of the two elements related to each other by this relation is the abbreviation of the other element.
- *Syntactic* : indicates that the two elements related to each other by this relation are syntactically equivalent, i.e. identical, or are acronyms, or one the two elements related to each other by this relation is syntactically included in the other element, or one of the two elements related to each other by this relation is the abbreviation of the other element.
- *Synonymy:* indicates that the two elements related to each other by this relation are synonyms
- *Disjunction:* indicates that the two elements (attributes) related to each other by this relation are disjoint.
- *Inverse:* indicates that the two elements (relations) related to each other by this relation are inverse.
- *EquivOnto:* indicates that the two elements (classes & relations) related to each other by this relation are equivalent (according to an ontology).
- *Semantic:* indicates that the two elements related to each other by this relation are semantically equivalent.
- *Global:* indicates that the two elements related to each other by this relation are equivalent in global structure.
- *Local:* indicates that the two elements related to each other by this relation are equivalent in local structure.
- *Structural:* indicates that the two elements related to each other by this relation are equivalent in global or local structure.
- *EquivalenceLevel: is* the level of equivalence between two elements. It depends on the equivalence links, i.e. it changes as a function of the equivalence links between elements. These levels help the user decide to validate or delete the mapping.
- *Sure:* indicates that the two elements related to each other by this relation are sure equivalent.
- *ModeratelySure:* indicates that the two elements related to each other by this relation are moderately sure equivalent.
- *Improbable:* indicates that the two elements related to each other by this relation are not very probable equivalent.

## 6. Case study

For illustration purposes, we present two simplified models, M1 and M2 (Figure 6 and 7), separately modeling the domain of a bank. M1 modeled an ATM (automated teller machine) and its relationship with customers, while M2 represents person's accounts in the bank. Only elements concerning our example are enrolled in models M1 and M2.

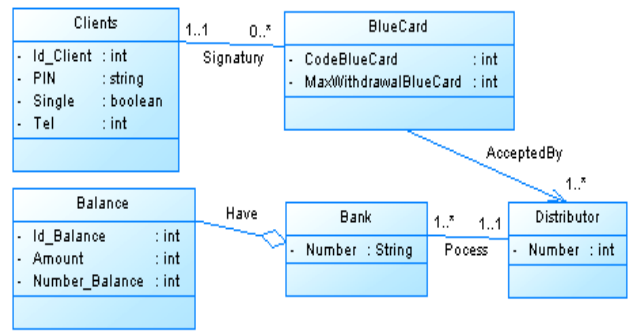

Fig.6. M1 model

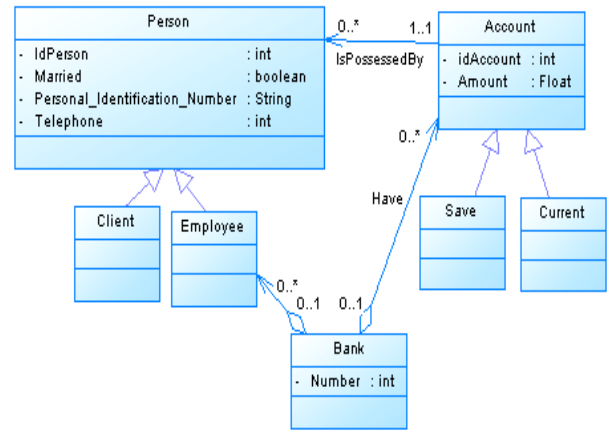

Fig.7. M2 model

To compare these models, we use the ontology of banking that we have modeled with Protégé [1] (Figure 8). Only elements that are relevant for our example are enrolled in this ontology. The ontology called "B" (Bank) contains the class "Person" and its subclasses "Client"; and "Employee". The "Person" class is characterized by the attributes (data properties) "PersonId", "Address", "Telephone" and "PersonalIdentificationNumber". The ontology also contains the classes "Bank" (characterized by "BankNumber"), "Distributor" (characterized by "DistibutorId"), "Account" (characterized by "AccountId" and "Amount") and its subclasses "Save", "Current" and "Mixed", classes "Balance", "budget", "Situation" and its subclasses "Single" and "Married"; and finally "BlueCard"; (characterized by "CodeBlueCard"; and "MaxWithdrawalBlueCard". "Account" is equivalent to "Balance" and "budget". "Single" and "Married" are disjoint. "Possess" is a relationship (object property) between "Person" and "Account". "IsPossessedBy" is the inverse of" Possess". The relationship "HaveSituation" links "Person" and "Situation".

---
[1] http://protege.stanford.edu/

The relationship "Accept" links 'Distributor' and "BlueCard". The relationship "Have" links "Bank" and "Distributor". Finally, the relationship "administer" links "Bank" and "Account". We also use in this example information from an abbreviations dictionary (Tel, Telephone) and an acronyms dictionary (PIN, Personal Identification Number).

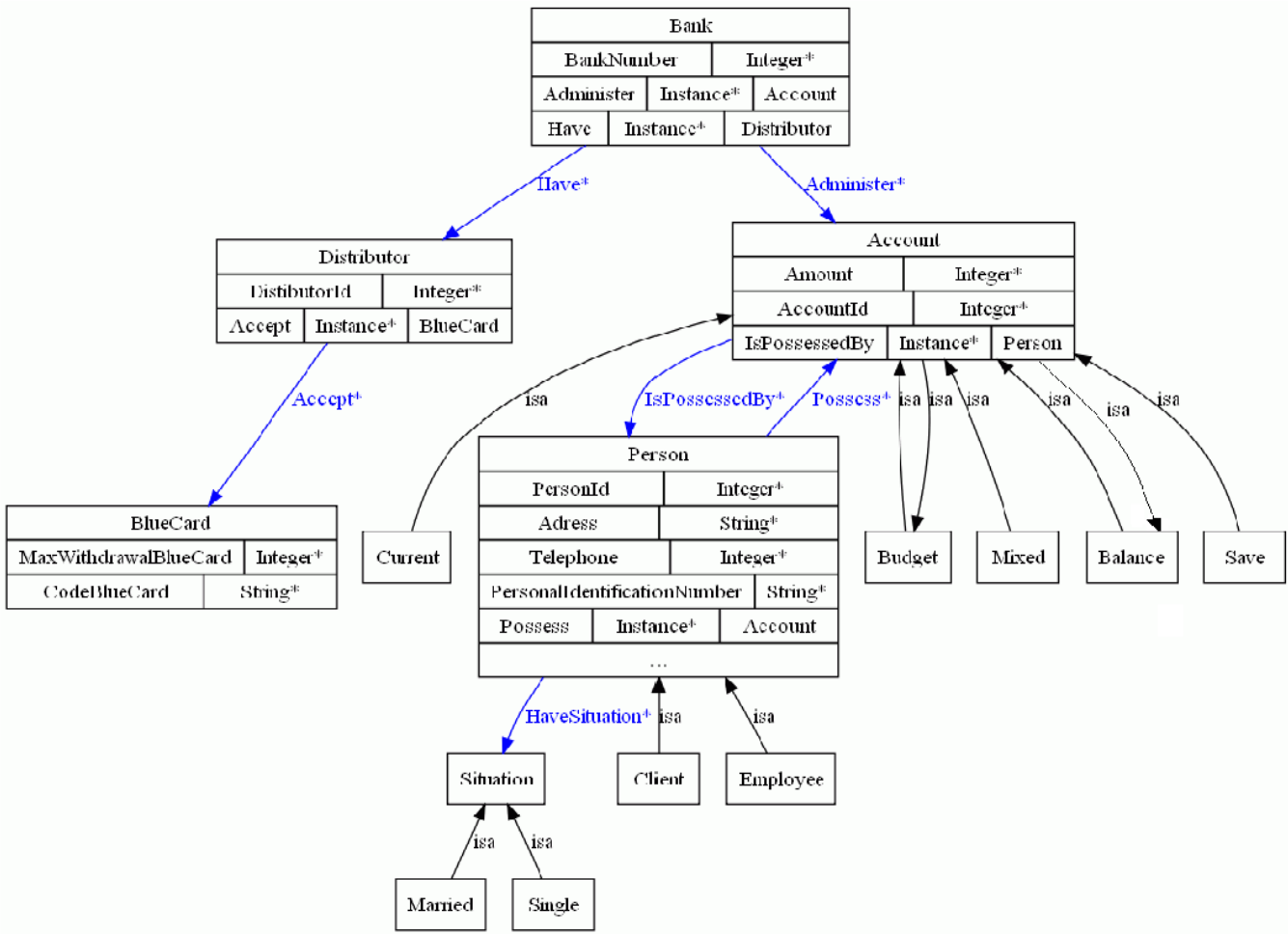

Fig.8. Ontology of banking domain

By applying the correspondence rules presented in [26], we obtain the correspondence model (figure 9, 10 and 11) conforming to our correspondence metamodel.

The first and second column represent respectively elements of M1 and M2. The third column represents syntactic or semantic equivalence. The fourth column represents the type of the previous column. The fifth and sixth columns represent respectively the equivalent in local and global structure. The seventh column represents the level of equivalence of elements. The last two columns (on the right) display buttons to validate or to delete a correspondence create automatically.

Fig. 9. Sure mapping

Fig.10. Moderately sure mapping

Fig. 11 Improbable mapping

Some elements of M1 and M2 have no correspondence; they must appear in the result model. Once the mapping is generated, the next step is integration.

## 7. Conclusion

Any approach for models comparison must take into account syntactic, semantic and structural aspects. The semantic comparison of models is a complex task because it requires understanding the semantics of linking concepts. The main contribution of this paper concerns the first step of the integration process with the syntactic, semantic and structural comparison of two models. We can schematize our approach by a continuity of points in Figure 1.

Our approach is a models transformation that has two models in input and produces a correspondence model conforming to correspondence metamodel described in this article. We expressed the transformation rules as correspondence rules in [26].

To implement this transformation, we express the UML metamodel and the correspondence metamodel in ECORE model. Correspondence rules are written with the ATL transformation language.

The integration can be applied to "n" models (M1 to Mn), integrating M1 and M2, and next, integrating their result model MR1/2 with M3, etc. The goal of our future research is to define integration and merging rules that will make possible the entire process of model integration.

**Samia Benabdellah** Master degree in Computer Science in 2008; PhD student at Mohammed V Souissi University, ENSIAS, Rabat, Morocco. Ongoing research interests: models Integration.

**Mounia Fredj** Professor at the software engineering department at ENSIAS; 10 recent publications papers between 2008 and 2010; Ongoing research interests: Information System engineering, components, patterns, MDA approach.

**Salma Mouline** Professor at the software engineering department at Faculty of Sciences, University Department, Mohammed V Souissi University. Ongoing research interests: Information System engineering, models.